\documentclass[twocolumn,aps,pra,superscriptaddress]{revtex4-1}
\usepackage{graphicx}
\usepackage{dcolumn}
\usepackage{bm}
\usepackage{multirow}
\usepackage{amssymb}
\usepackage{amsmath}
\usepackage{graphicx}
\usepackage{xcolor}
\usepackage[colorlinks,citecolor=blue,linkcolor=red,urlcolor=blue]{hyperref}
\usepackage[all]{hypcap}
\usepackage{indentfirst}
\usepackage{amsmath}
\usepackage{cases}

\begin{document}
\title{Kibble-Zurek scaling in one-dimensional localization transitions}
\author{Xuan Bu\footnotemark[1]}
\altaffiliation{These authors contribute equally to this work.}
\affiliation{School of Physics, Sun Yat-Sen University, Guangzhou 510275, China}
\author{Liang-Jun Zhai\footnotemark[1]}
\altaffiliation{These authors contribute equally to this work.}
\affiliation{The school of mathematics and physics, Jiangsu University of Technology, Changzhou 213001, China}
\affiliation{Department of Physics, Nanjing University, Nanjing 210093, China}
\author{Shuai Yin}
\email{yinsh6@mail.sysu.edu.cn}
\affiliation{School of Physics, Sun Yat-Sen University, Guangzhou 510275, China}
\date{\today}
\begin{abstract}
In this work, we explore the driven dynamics of the one-dimensional ($1$D) localization transitions. By linearly changing the strength of disorder potential, we calculate the evolution of the localization length $\xi$ and the inverse participation ratio (IPR) in a disordered Aubry-Andr\'{e} (AA) model, and investigate the dependence of these quantities on the driving rate. At first, we focus on the limit in the absence of the quasiperiodic potential. We find that the driven dynamics from both ground state and excited state can be described by the Kibble-Zurek scaling (KZS). Then, the driven dynamics near the critical point of the AA model is studied. Here, since both the disorder and the quasiperiodic potential are relevant directions, the KZS should include both scaling variables. Our present work not only extends our understanding of the localization transitions but also generalize the application of the KZS.
\end{abstract}
\footnotetext[1]{These authors contribute equally to this work.}
\maketitle

\section{Introduction}
The physics of phase transitions between localized and metallic phases in disordered systems have attracted long-term attentions since the pioneering work of Anderson~\cite{Anderson1958,Thouless1974,Evers2008,Hatano1998,Hamizaki2019,Abanin2019}.
As a result of the destructive interference of scattered waves, the wave function can be localized at some isolated sites.
Theoretically, it was shown that for one- and two-dimensional disordered systems, the localization transition happens for infinitesimal disorder strength, whereas for higher-dimensional systems, the localization transition happens for finite disorder strength~\cite{Thouless1974,Evers2008}.
Moreover, universality classes of Anderson transition have been categorized~\cite{Shinobu1981,Alexander1997,TWang2021,XLuo2021,XLuo2022}.
In addition, besides the disordered systems, it was shown that the localization can also happen in quasiperiodic systems~\cite{Aubry1980,Sokoloff1981,Sarma1988,Biddle2009,Biddle2011,Luschen2018,Skipetrov2018,Agrawal2020,Zeng2017,Longhi2019,Tang2020,Jiang2019,zhai2020,zhai2021,Liu20212,Sahoo2022,Jazaeri2001,Kawabata2021, Wang2021,Strkalj2021,huseprl,zhangsx2018}.
For instance, it was shown that the Aubry-Andr\'{e} (AA) model hosts a localization transition at finite strength of quasiperiodic potential~\cite{Aubry1980,Sokoloff1981,Zeng2017,Longhi2019,Tang2020,Jiang2019,zhai2020,zhai2021,Liu20212}.
Experimentally, the localization transition has been observed in various platforms~\cite{Semeghini2015,DHWhite2020,Wiersma1997,Mookherjea2014,HFHu2008,Lee1985}, such as cold atomic systems~\cite{Semeghini2015,DHWhite2020}, quantum optics~\cite{Wiersma1997,Mookherjea2014}, acoustic waves~\cite{HFHu2008}, and electronic systems~\cite{Lee1985}.

On the other hand, great progresses have been made in controlling quantum matter with high precision in the last decades, inspiring the investigations on the nonequilibrium dynamics of quantum systems~\cite{Gross2017,Blatt2012,Mukherjee2017,Xiao2021}.
In particular, the driven dynamics across a critical point has aroused wide concern due to its potential application in adiabatic quantum computations~\cite{Reichhardt2022}.
A general theory describing the driven critical dynamics is the celebrated Kibble-Zurek scaling (KZS)~\cite{Laguna1997,Yates1998,Dziarmaga2005,Polkovnikov2011,Das2012,Adolfo2014,Feng2016,Yin2017,zhai2018}.
By linearly changing the distance to the critical point, the KZS states that the whole driven process can be divided into different stages.
In the initial stage, the system evolves adiabatically along the equilibrium state.
Then, the system enters an impulse region, in which the evolution of the system lags behind the external driving as a result of the critical slowing down.
A full finite-time scaling form with the driving rate being a typical scaling variable has been proposed in characterizing the nonequilibrium dynamics in the whole process~\cite{Zhong2006,Gong2010,Huang2014}.
This full scaling form has been verified in both classical and quantum phase transitions~\cite{Monaco2002,Anglin1999,Antunes2006,Dziarmaga2005,Damski2007,Chandran2012}.

Recently, the nonequilibrium dynamics in the localization transition have also attracted increasing attentions,
which have extended our understanding of localization transitions and universality far from equilibrium~\cite{Molina2014,Bairey2014,Yang2017,Romito2018,Yin2018,Decker2020,Modak2021,Xu2021,Sinha2019,Zhai2022,zhai2022b}.
For instance, in disordered systems, dynamical phase transition characterized by the peaks in the Loschmidt echo after a sudden quench was studied~\cite{Yin2018}.
In addition, the KZS has been investigated in the localization transitions in quasiperiodic AA model and its non-Hermitian variant for changing the quasiperiodic potential to cross the critical point~\cite{Sinha2019,Zhai2022,zhai2022b}.
However, there is still unknown whether the KZS is applicable for changing the disorder strength.

In this work, we study the driven dynamics of localization transitions in one-dimensional (1D) disordered systems.
We illustrate the dynamic scaling in a disordered AA model and focus on two cases.
In the first case, there is no quasiperiodic potential and this model recovers the usual Anderson model.
In the second case, the system is located near the AA critical point. For both cases, we change the disorder coefficient across the transition point and calculate the evolution of the localization length $\xi$ and the inverse participation ratio (IPR). For the Anderson model, we find that the evolution of these quantities satisfy the usual KZS from both ground state and highest excited state; whereas for the disordered AA model, since the quasiperiodic potential is another relevant direction, the full scaling form should also include the contribution from this term. In particular, in the overlap region between the critical regions of the AA model and the Anderson transition, we show that the dynamic scaling behaviors can be described by both the AA critical exponents and the critical exponents of the Anderson localization.

The rest of the paper is arranged as follows.
The $1$D disordered AA model and the characteristic quantities are introduced in Sec.~\ref{secmodel}. In Sec.~\ref{kzanderson}, the driven dynamics in the Anderson model is studied. Then, we explore the driven dynamics near the AA critical point in Sec.~\ref{kzdaa}.
A summary is given in Sec.~\ref{secSum}.

\section{\label{secmodel}Model and static scaling properties}
The Hamiltonian of the disordered AA model reads~\cite{Bu2022}
\begin{eqnarray}
\label{Eq:model}
H &=& -J\sum_{j}^{L}{(c_{j}^\dagger c_{j+1}+h.c.})\\ \nonumber
&&+(2J+\delta)\sum_{j}^{L}\cos{[2\pi(\gamma j+\phi)]c_j^\dagger c_j}\\ \nonumber
&&+\varepsilon \sum_{j}^{L}w_j c_j^\dagger c_j,
\end{eqnarray}
in which $c_j^\dagger (c_j)$ is the creation (annihilation) operator of the hard-core boson at site $j$,
and $J$ is the hopping amplitude between the nearest-neighboring sites and is chosen as the unity of energy, $(2J+\delta)$ measures the amplitude of the quasiperiodic potential, $\gamma$ is an irrational number, $\phi$ is the phase of the potential with a uniform distribution in $[0,1)$, $w_j$ provides the quenched disorder distributed uniformly in the interval of $[-1,1]$,
and $\varepsilon$ is the coefficient of the disorder.
To satisfy the periodic boundary condition, $\gamma$ has to be approximated by a rational number $F_n/F_{n+1}$ where $F_{n+1}=L$ and $F_{n}$ are the Fibonacci numbers~\cite{Jiang2019,Zhai2022}.

\begin{figure}[tbp]
\centering
  \includegraphics[width=0.7\linewidth,clip]{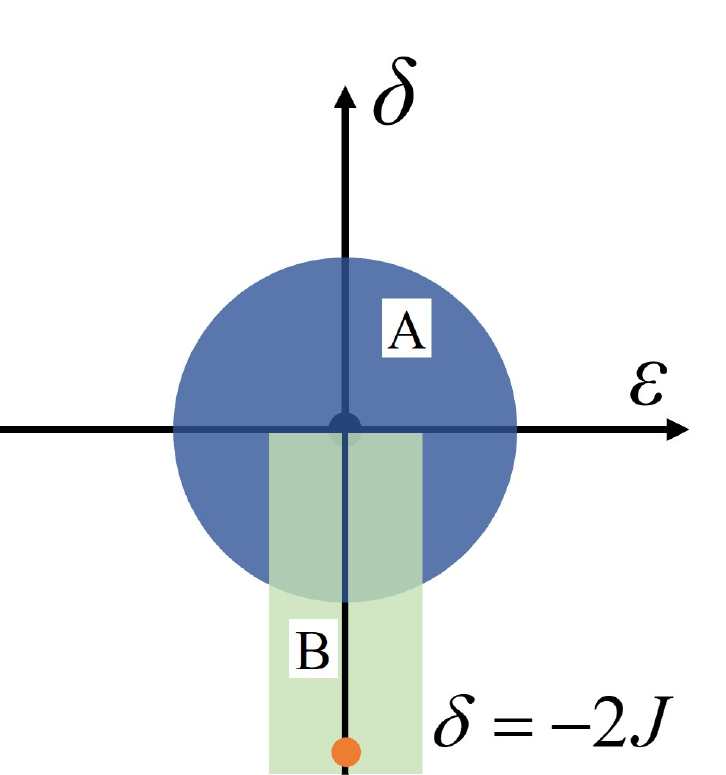}
  \vskip-3mm
  \caption{Sketch of the phase diagram of the disorder AA model. When $\delta=-2J$ (denoted by the yellow point), this model recovers the Anderson model. The dark blue region (region A) denotes the critical region of localization transition of the disordered AA model.
  The light green region (region B) denotes the critical region of the Anderson localization transition.
  Near the critical point of $\delta=0$ and $\varepsilon=0$, these critical regions overlap with each other.
  }
  \label{phase}
\end{figure}

The phase diagram of model~(\ref{Eq:model}) is shown in Fig.~\ref{phase}. For $\delta=-2J$, Eq.~(\ref{Eq:model}) recovers the Anderson model in which all states are localized for any finite $\varepsilon$. Thus its critical point of the localization transition is at $\varepsilon=0$. In the critical region, the localization length $\xi$, defined as~\cite{Sinha2019,Zhai2022,Bu2022}
\begin{equation}
\label{Eq:xiscaling}
   \xi = \sqrt{\sum_{j}^{L} [( j - j_c )^2 ] P_j},
\end{equation}
with $P_j$ being the probability of the wave function at site $j$, and $j_c\equiv\sum_j jP_j$ being the localization center, diverges as
\begin{equation}
\label{Eq:xiscaling1}
\xi\propto \varepsilon^{-\nu},
\end{equation}
in which $\nu=2/3$~\cite{Wei2019,Bu2022}. Another quantity to characterize the localization transition is the inverse participation ratio (IPR), which is defined as~\cite{Bauer1990,Fyodorov1992}
\begin{equation}
\label{Eq:ipr}
{\rm IPR} = \sum_{j=1}^L|\Psi(j)|^4,
\end{equation}
where $\Psi(j)$ is the wavefunction. For a localized state, the wave function is localized on some isolated sites, and ${\rm IPR}\propto L^0$, whereas ${\rm IPR}\propto L^{-1}$ for the delocalized states. Close to the critical point, ${\rm IPR}$ scales with $\varepsilon$ as
\begin{equation}
\label{Eq:iprscaling1}
{\rm IPR}\propto \varepsilon^{s},
\end{equation}
with the critical exponent being $s=2/3$~\cite{Bu2022}. In addition, the dynamic exponent $z$ for the Anderson model is $z=2$~\cite{Wei2019}.

For $\varepsilon=0$, Eq.~(\ref{Eq:model}) recovers the AA model. It was shown that all the eigenstates of are localized when $\delta>0$, and all the eigenstates are delocalized when $\delta<0$. In the critical region, the localization length $\xi$ satisfies
\begin{equation}
\label{Eq:xiscaling2}
\xi\propto \delta^{-\nu_\delta},
\end{equation}
with $\nu_\delta=1$~\cite{Sinha2019,Wei2019}. And the IPR obeys
\begin{equation}
\label{Eq:iprscaling2}
{\rm IPR}\propto \delta^{s_\delta},
\end{equation}
with $s_\delta\approx 0.333$~\cite{Bu2022}. Besides, the dynamic exponent $z$ for the AA critical point is $z_\delta\approx 2.37$~\cite{Sinha2019}.

Moreover, previously we showed that the disorder $\varepsilon$ also provides a relevant direction in the AA critical point. For $\delta=0$, the localization length $\xi$ obeys
\begin{equation}
\label{Eq:xiscaling3}
\xi\propto \varepsilon^{-\nu_\varepsilon},
\end{equation}
with $\nu_\varepsilon=0.46(1)$~\cite{Bu2022}.
Note that this exponent is remarkably different from $\nu$ and $\nu_\delta$. In addition, the IPR obeys
\begin{equation}
\label{Eq:iprscaling3}
{\rm IPR}\propto \varepsilon^{s_\delta \nu_\varepsilon/\nu_\delta}.
\end{equation}

\section{\label{secdynamics} The KZS in the localization transition}

\subsection{\label{kzanderson} KZS for the Anderson model}
Here, we consider the driven dynamics of the Anderson model with $\delta=-2J$ in Eq.~(\ref{Eq:model}).
At first, we show the detailed driven process.
Initially, the system is in the localization phase for a specific realization of $w_j$ with coefficient $\varepsilon_0>0$.
Then $\varepsilon$ is decreased according to
\begin{eqnarray}
  \varepsilon &=&\varepsilon_0-Rt,
\end{eqnarray}
to cross the critical point, and $w_j$ keep invariant. Then $w_i$ is resampled for another process with same initial $\varepsilon_0$. At last, the quantities are averaged for many realization of samples to make the evolution curves smooth.

The KZS states that when $\varepsilon>R^{1/\nu r}$ with $r=z+1/\nu$, the system can evolve adiabatically since the state has enough time to adjust to the change in the Hamiltonian;
in contrast, when $\varepsilon<R^{1/\nu r}$, the system enter the impulse region and the system stop evolving as a result of the critical slowing down.
However, investigations showed that the assumption that the system does not evolve in the impulse region is oversimplified.
To improve it, a finite-time scaling theory has been proposed and demonstrates that the external driving provides a typical time scale of $\zeta\propto R^{-z/r}$~\cite{Zhong2006,Gong2010,Huang2014}.
In the impulse region, $\zeta$ controls the dynamic scaling behaviors and macroscopic quantities can be scaled with $\zeta$.
For instance, for large enough system size, the full scaling form of the localization length $\xi$ around the critical point reads~\cite{Zhai2022}
\begin{eqnarray}
 \label{Eq:dynamicScalingXiR}
 \xi(\varepsilon,R) &=& R^{-1/r}f_1(\varepsilon R^{-1/r\nu}),
\end{eqnarray}
in which $f_1$ is the scaling function. When $\varepsilon> R^{-1/r\nu}$, the evolution is in the adiabatic stage, in which $f_1(\varepsilon R^{-1/r\nu})\sim(\varepsilon R^{-1/r\nu})^{-\nu}$. Accordingly, $\xi$ satisfies Eq.~(\ref{Eq:xiscaling1}) and does not depend on the driving rate $R$. In contrast, near the critical point, when $\varepsilon< R^{-1/r\nu}$, $f_1(\varepsilon R^{-1/r\nu})$ tends to a constant and $\xi\propto R^{-1/r}$, demonstrating that the divergence of $\xi$ at the critical point has been truncated by the external driving and $\xi$ decreases as $R$ increases.

Similarly, the driven dynamics of the IPR around the critical point satisfies
\begin{equation}
 \label{Eq:dynamicScalingipr}
 {\rm IPR}(\varepsilon,R)=R^{s/r \nu}f_{2}(\varepsilon R^{-1/r\nu}).
\end{equation}
When $\varepsilon> R^{-1/r\nu}$, $f_{2}(\varepsilon R^{-1/r\nu})\sim (\varepsilon R^{-1/r\nu})^s$ and Eq.~(\ref{Eq:dynamicScalingipr}) recovers Eq.~(\ref{Eq:iprscaling1}). In contrast, near the critical point, when $\varepsilon< R^{-1/r\nu}$, ${\rm IPR}\propto R^{s/r}$.

\begin{figure}[tbp]
\centering
  \includegraphics[width=\linewidth,clip]{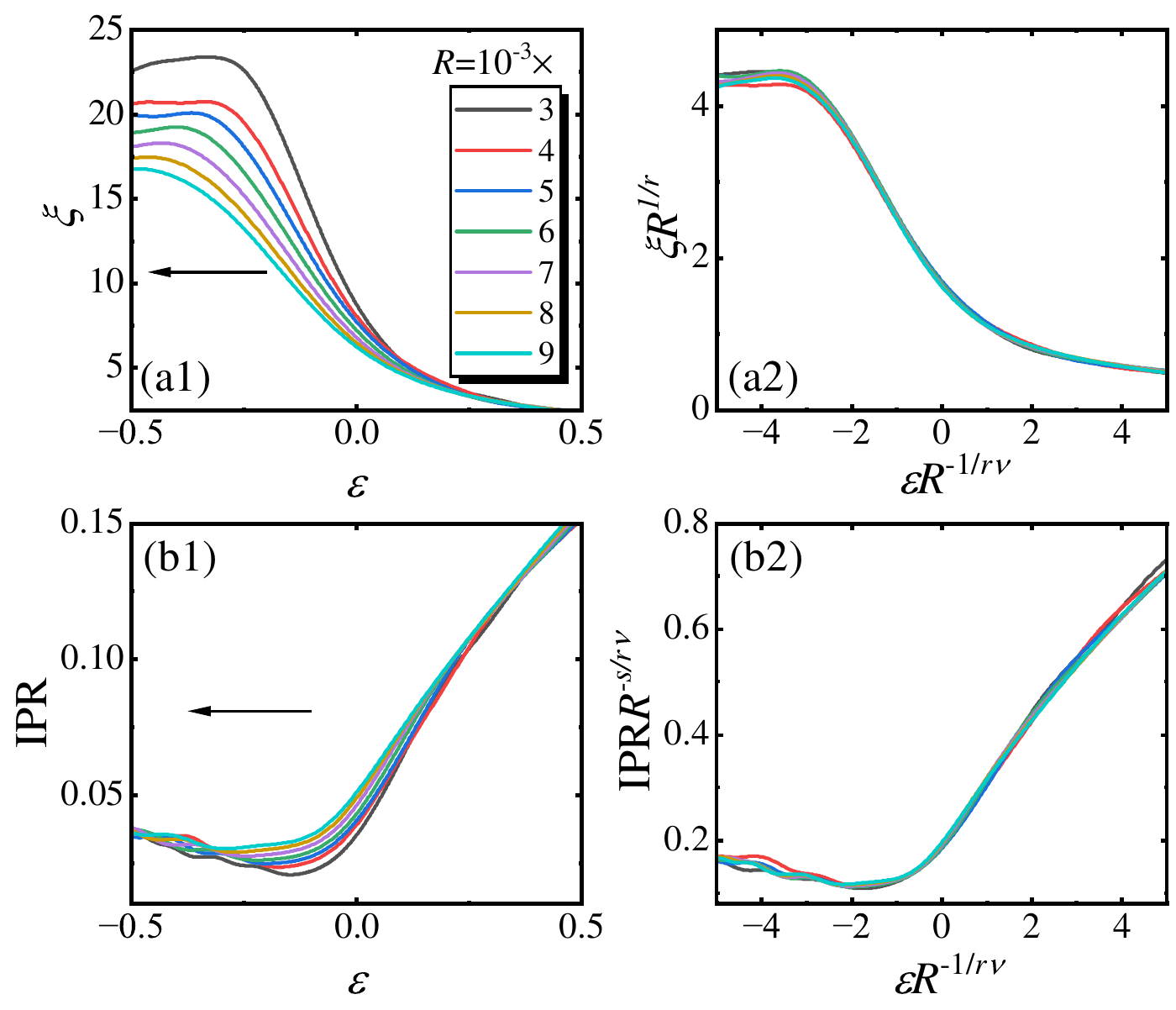}
  \vskip-3mm
  \caption{Driven dynamics in the Anderson model with the initial state being the ground state.
  The curves of $\xi$ versus $\varepsilon$ before (a1) and after (a2) rescaled for different $R$.
  The curves of ${\rm IPR}$ versus $\varepsilon$ before (b1) and after (b2) rescaled for different $R$.
  The arrows in (a1) and (b1) point the quench direction.
  }
  \label{dynground}
\end{figure}
\begin{figure}[tbp]
\centering
  \includegraphics[width=\linewidth,clip]{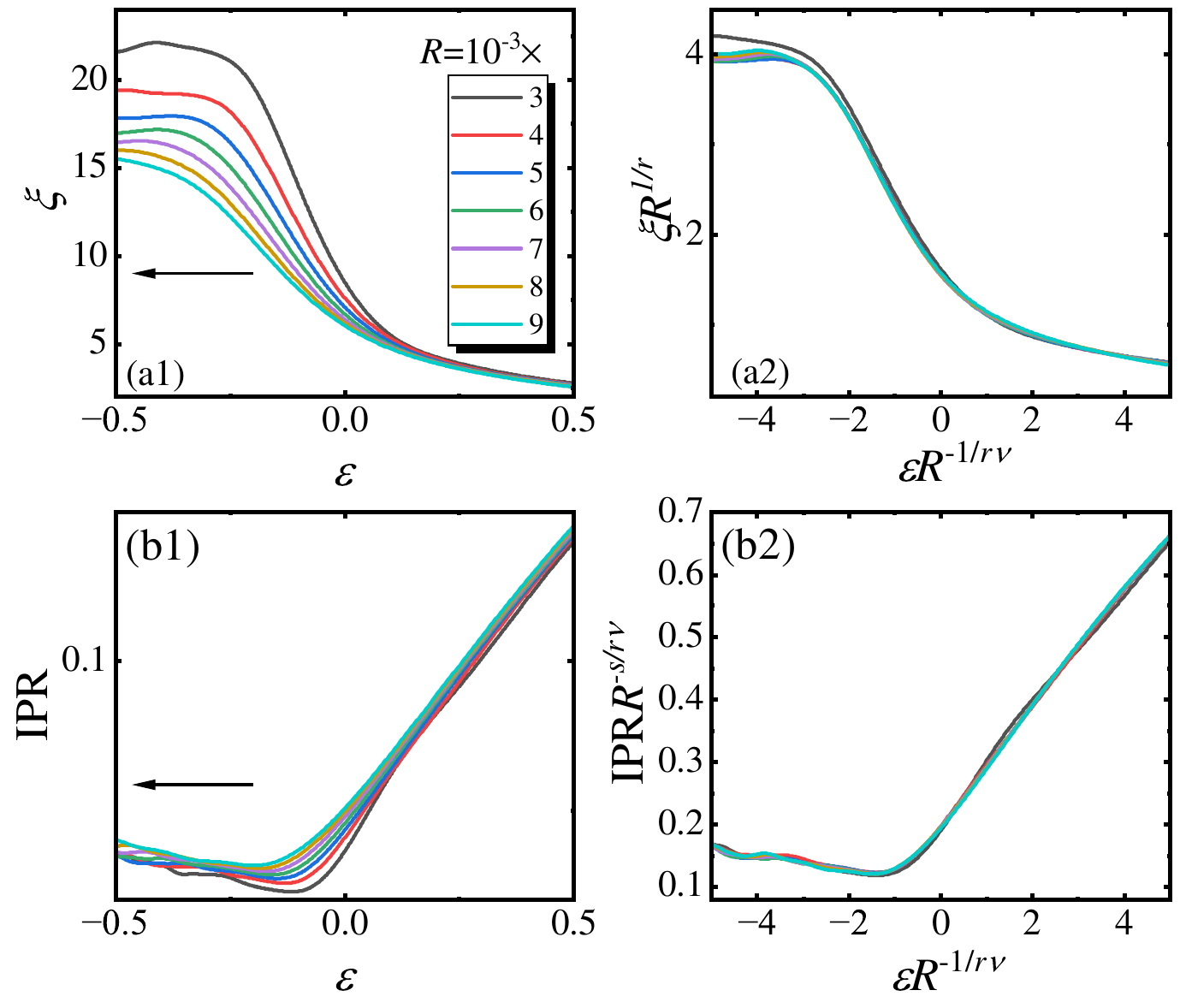}
  \vskip-3mm
  \caption{Driven dynamics in the Anderson model with the initial state being the highest excited state.
  The curves of $\xi$ versus $\varepsilon$ before (a1) and after (a2) rescaled for different $R$.
  The curves of ${\rm IPR}$ versus $\varepsilon$ before (b1) and after (b2) rescaled for different $R$.
  The arrows in (a1) and (b1) point the quench direction.
  }
  \label{dynexcited}
\end{figure}

To verify the scaling functions of Eq.~(\ref{Eq:dynamicScalingXiR}) and (\ref{Eq:dynamicScalingipr}), we numerically solve the Schrodinger equation for model~(\ref{Eq:model}), and calculate the dependence of $\xi$ and ${\rm IPR}$ on $\varepsilon$ for various driving rate $R$.
The finite difference method in the time direction is used, and the time interval is chosen as $10^{-3}$.
The lattice size is chosen as $L=500$, which is large enough to ignore the finite-size effect.
$\varepsilon_0$ is set as $\varepsilon_0=2$, which is far enough from the critical point at $\varepsilon=0$.

First, the initial state is chosen as the ground state of model~(\ref{Eq:model}) for $\varepsilon=\varepsilon_0$. Figure~\ref{dynground} (a1) shows the evolution of the localization length $\xi$ for different $R$. Initially, one finds that $\xi$ almost does not depend on $R$, indicating the system evolves adiabatically in this stage. Then when $\varepsilon$ approaches to the critical point, the curves for different $R$ begin to separate from each other, indicating that the system enters the impulse region. After rescaling $\xi$ and $\varepsilon$ as $\xi R^{1/r}$ and $\varepsilon R^{-1/\nu r}$, respectively, we find that the rescaled curves collapse onto each other near the critical point, as shown in Fig.~~\ref{dynground} (a2). These results confirm Eq.~(\ref{Eq:dynamicScalingXiR}). In particular, exactly at the critical point, i.e., $\varepsilon=0$, Fig.~~\ref{dynground} (a2) demonstrates $\xi\propto R^{-1/r}$.

Similarly, Fig.~\ref{dynground} (b1) shows the evolution of IPR for different $R$. After an initial adiabatic stage, in which the evolution of IPR is almost independent of $R$, hysteresis effect of IPR appears near the critical point and the IPR increases as $R$ increases. After rescaling IPR and $\varepsilon$ as ${\rm IPR} R^{-s/\nu r}$ and $\varepsilon R^{-1/\nu r}$, respectively, we find that the rescaled curves match with each other near the critical point, as shown in Fig.~~\ref{dynground} (b2). These results confirm Eq.~(\ref{Eq:dynamicScalingipr}). In particular, exactly at the critical point, i.e., $\varepsilon=0$, Fig.~\ref{dynground} (b2) demonstrates ${\rm IPR} \propto R^{s/\nu r}$. These results clearly demonstrates that the KZS is applicable in the localization transition of the Anderson model.

Moreover, different from the usual quantum phase transition which happens only in the ground state, here the Anderson localization happens in all eigenstates. It is interesting to explore the driven dynamics with the initial state being the excited state. To this end, we calculate the dynamics of $\xi$ and IPR with the initial state being the highest excited state and show the results in Fig.~\ref{dynexcited}. After rescaling the curves by $R$, we find that the rescaled curves collapse onto each other, as shown in Fig.~\ref{dynexcited}, verifying Eqs.~(\ref{Eq:dynamicScalingXiR}) and (\ref{Eq:dynamicScalingipr}) and demonstrating that the KZS is still applicable in the driven dynamics from the excited states.

\subsection{\label{kzdaa}KZS for the disordered AA model}
\begin{figure}[tbp]
\centering
  \includegraphics[width=\linewidth,clip]{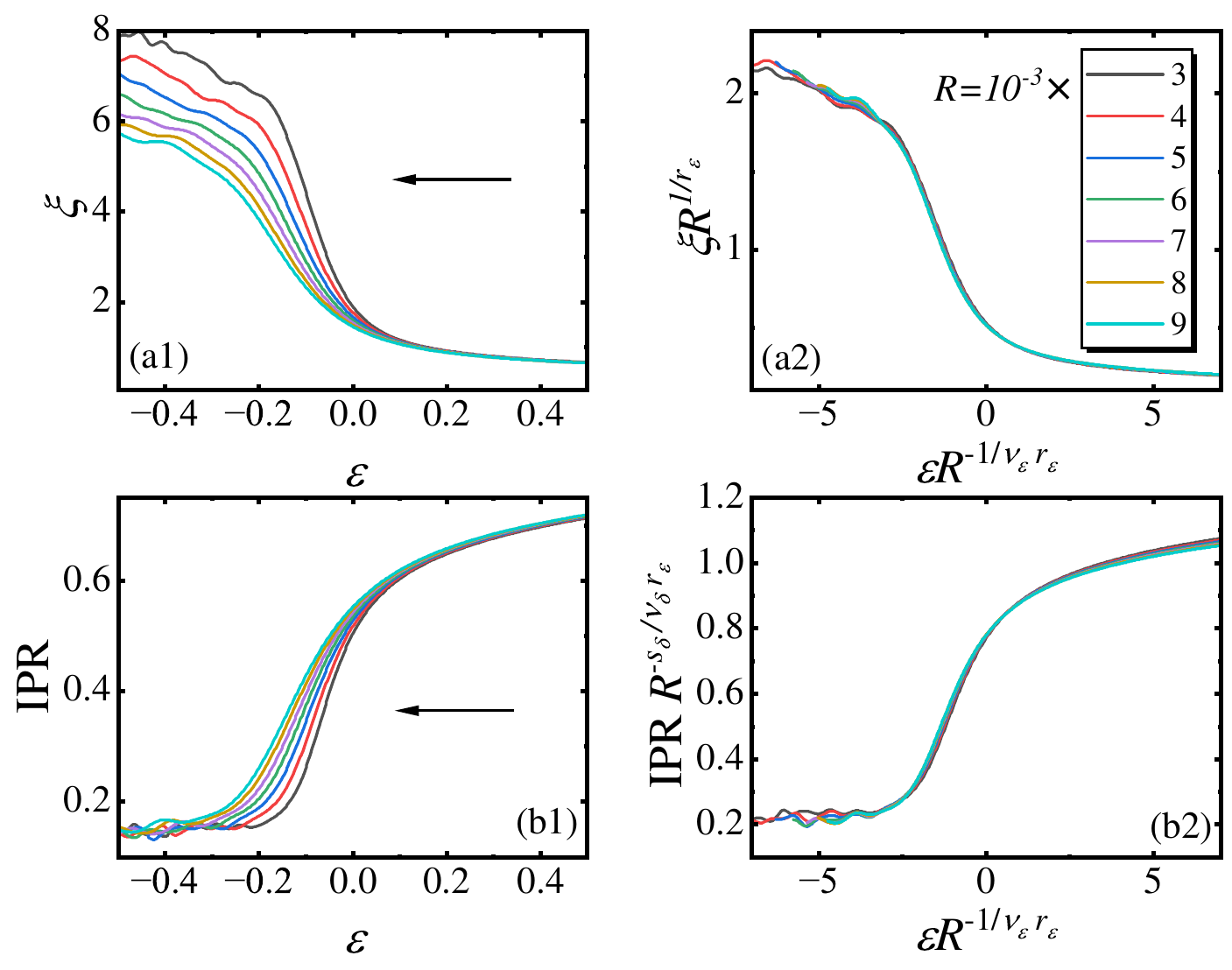}
  \vskip-3mm
  \caption{Driven dynamics near the AA critical point with fixed $\delta R^{-1/r_\varepsilon\nu_\delta}=0.3$.
  The curves of $\xi$ versus $\varepsilon$ before (a1) and after (a2) rescaled for different $R$.
  The curves of ${\rm IPR}$ versus $\varepsilon$ before (b1) and after (b2) rescaled for different $R$.
  The arrows in (a1) and (b1) point the quench direction.
  }
  \label{dyngroundaa1}
\end{figure}

\begin{figure}[tbp]
\centering
  \includegraphics[width=\linewidth,clip]{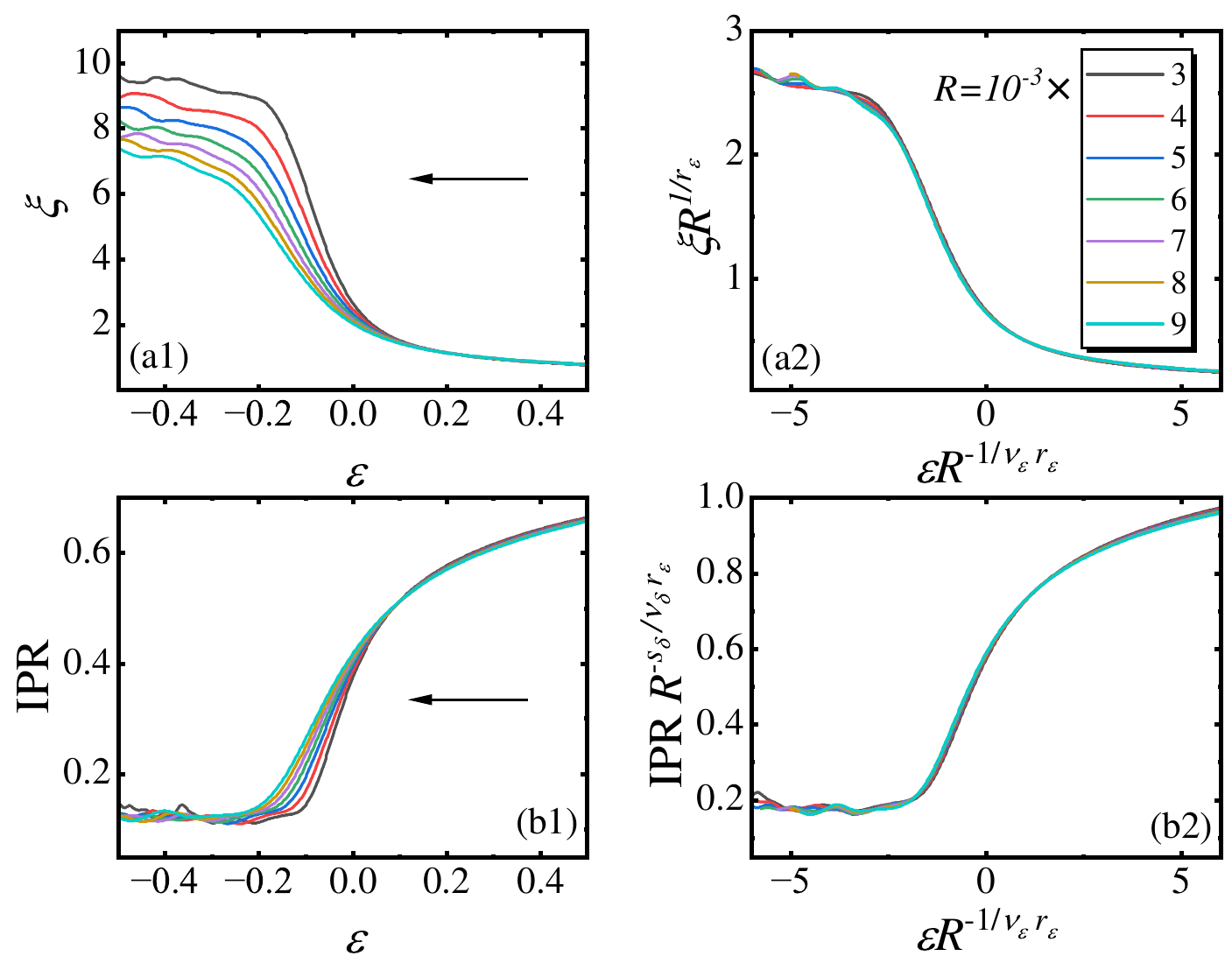}
  \vskip-3mm
  \caption{Driven dynamics near the AA critical point with fixed $\delta R^{-1/r_\varepsilon\nu_\delta}=-0.3$.
  The curves of $\xi$ versus $\varepsilon$ before (a1) and after (a2) rescaled for different $R$.
  The curves of ${\rm IPR}$ versus $\varepsilon$ before (b1) and after (b2) rescaled for different $R$.
  The arrows in (a1) and (b1) point the quench direction.
  }
  \label{dyngroundaa2}
\end{figure}

\begin{figure}[tbp]
\centering
  \includegraphics[width=\linewidth,clip]{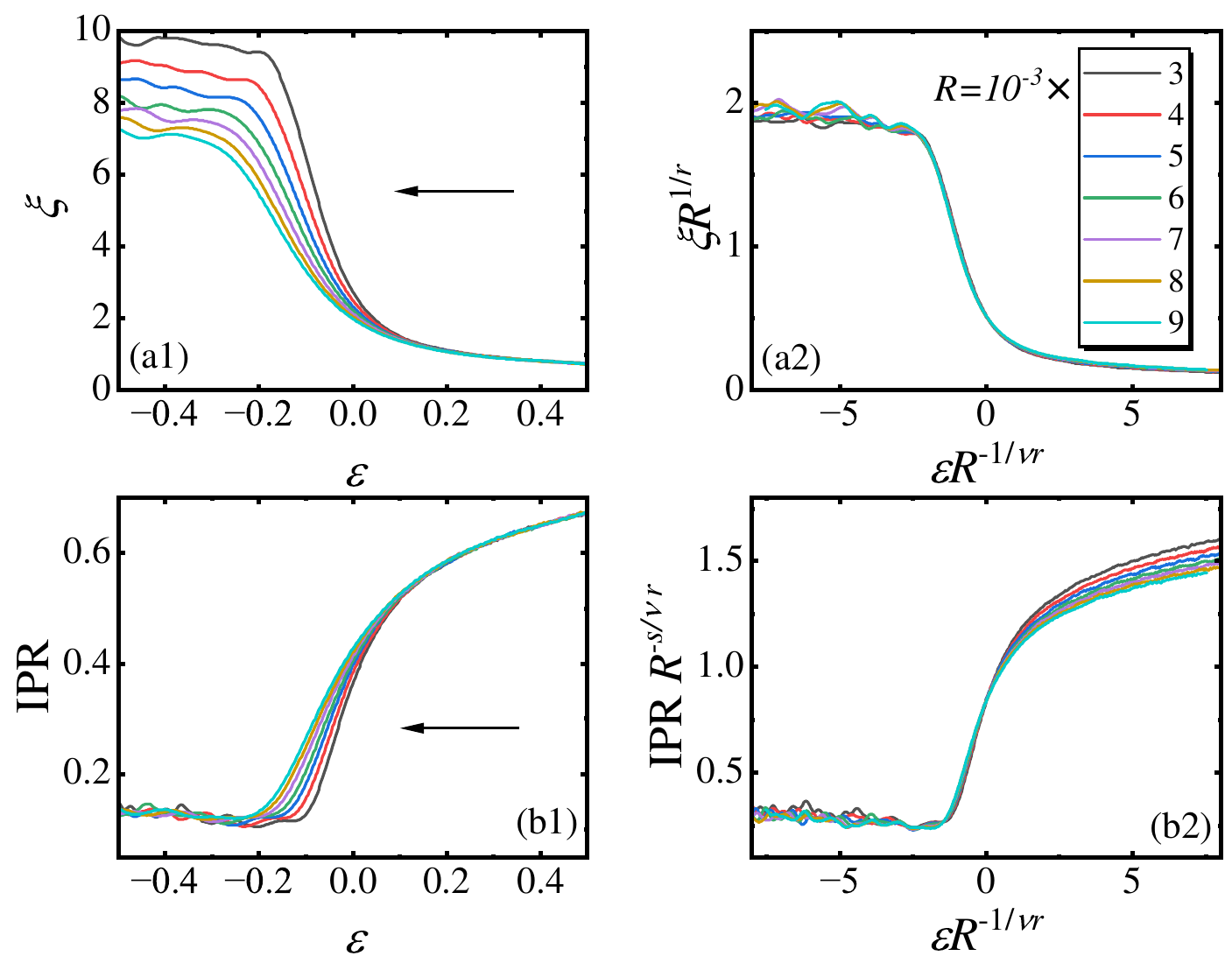}
  \vskip-3mm
  \caption{Driven dynamics near the AA critical point with fixed $\delta=-0.1$.
  The curves of $\xi$ versus $\varepsilon$ before (a1) and after (a2) rescaled for different $R$.
  The curves of ${\rm IPR}$ versus $\varepsilon$ before (b1) and after (b2) rescaled for different $R$.
  The arrows in (a1) and (b1) point the quench direction.
  }
  \label{dyngroundaa3}
\end{figure}

In this section, we consider the driven dynamics near the AA critical point with small $\delta$ in model~(\ref{Eq:model}) by changing the coefficient of the disorder term. Note that different from the Anderson model, there are two relevant directions near the critical point of the disordered AA model. One direction is the quasiperiodic potential, represented by $\delta$, the other is the disorder term, represented by $\varepsilon$. Thus, in the full scaling form, both two relevant terms should be included.

In analogy to the analyses in Sec.~\ref{kzanderson}, the evolution of the localization length $\xi$ should satisfy
\begin{eqnarray}
 \label{Eq:dynamicScalingXiR1}
 \xi(\varepsilon,\delta,R) &=& R^{-1/r_\varepsilon}f_3(\varepsilon R^{-1/r_\varepsilon\nu_\varepsilon},\delta R^{-1/r_\varepsilon\nu_\delta}),
\end{eqnarray}
in which $r_\varepsilon=z_\delta+1/\nu_\varepsilon$. For $R\rightarrow 0$ and $\delta=0$, $f_3\sim (\varepsilon R^{-1/r_\varepsilon\nu_\varepsilon})^{-\nu_\varepsilon}$ and Eq.~(\ref{Eq:dynamicScalingXiR1}) restores Eq.~(\ref{Eq:xiscaling3}). For $R\rightarrow 0$ and $\varepsilon=0$, $f_3\sim (\delta R^{-1/r_\varepsilon\nu_\delta})^{-\nu_\delta}$ and Eq.~(\ref{Eq:dynamicScalingXiR1}) restores Eq.~(\ref{Eq:xiscaling2}).

Similarly, under external driving, the IPR should satisfy
\begin{eqnarray}
 \label{Eq:dynamicScalingipr1}
 {\rm IPR} (\varepsilon,\delta,R) &=& R^{s_\delta /r_\varepsilon \nu_\delta}f_4(\varepsilon R^{-1/r_\varepsilon\nu_\varepsilon},\delta R^{-1/r_\varepsilon\nu_\delta}).
\end{eqnarray}
For $R\rightarrow 0$ and $\delta=0$, $f_4\sim (\varepsilon R^{-1/r_\varepsilon\nu_\varepsilon})^{s_\delta \nu_\varepsilon/\nu_\delta}$ and Eq.~(\ref{Eq:dynamicScalingipr1}) recovers Eq.~(\ref{Eq:iprscaling3}). For $R\rightarrow 0$ and $\varepsilon=0$, $f_3\sim (\delta R^{-1/r_\varepsilon\nu_\delta})^{s_\delta}$ and Eq.~(\ref{Eq:dynamicScalingipr1}) restores Eq.~(\ref{Eq:iprscaling2}).

Equations~(\ref{Eq:dynamicScalingXiR1}) and (\ref{Eq:dynamicScalingipr1}) should be applicable for any values of $\delta$ and $\varepsilon$ near the critical point of the AA model. Particularly, for $\delta<0$, there is an overlap critical region between the critical region of the AA critical point and the critical region of the Anderson localization, as illustrated in Fig.~\ref{phase}. Therefore, in this overlap region, the driven critical dynamics of should simultaneously satisfy Eqs.~(\ref{Eq:dynamicScalingXiR}) and (\ref{Eq:dynamicScalingXiR1}) for $\xi$ and Eqs.~(\ref{Eq:dynamicScalingipr}) and (\ref{Eq:dynamicScalingipr1}) for the IPR.

We at first examine Eqs.~(\ref{Eq:dynamicScalingXiR1}) and (\ref{Eq:dynamicScalingipr1}) for $\delta>0$ with the initial state being the ground state. For a fixed $\delta R^{-1/r_\varepsilon\nu_\delta}$, we calculate the evolution of $\xi$ and IPR for various driving rate $R$. After rescaling the evolution curves with $R$, we find that the rescaled curves match with each other, as shown in Fig.~\ref{dyngroundaa1}, confirming Eqs.~(\ref{Eq:dynamicScalingXiR1}) and (\ref{Eq:dynamicScalingipr1}).

For $\delta<0$, Fig.~\ref{dyngroundaa2} shows the evolution of $\xi$ and IPR with various driving rate $R$ for a fixed $\delta R^{-1/r_\varepsilon\nu_\delta}$. After rescaling the curves by $R$ with the critical exponents of the AA critical point, we find that the curves collapse onto each other, as shown in Fig.~\ref{dyngroundaa2}, confirming Eqs.~(\ref{Eq:dynamicScalingXiR1}) and (\ref{Eq:dynamicScalingipr1}). Moreover, Fig.~\ref{dyngroundaa3} shows the evolution of $\xi$ and IPR for various driving rate $R$ with a fixed $\delta$, which is near the AA critical point. After rescaling the curves by $R$ with the critical exponents of the Anderson model, we find that the curves also collapse onto each other, as shown in Fig.~\ref{dyngroundaa3}, obeying Eqs.~(\ref{Eq:dynamicScalingXiR}) and (\ref{Eq:dynamicScalingipr}). Thus, we confirm that for $\delta<0$ the driven critical dynamics can simultaneously be described by Eqs.~(\ref{Eq:dynamicScalingXiR}) and (\ref{Eq:dynamicScalingXiR1}) for $\xi$ and Eqs.~(\ref{Eq:dynamicScalingipr}) and (\ref{Eq:dynamicScalingipr1}) for the IPR.

Here we remark on the results. (a) Although here we only show the results with the initial state being the ground state, it is expected that these scaling analyses are also applicable for the excited states, similar to the results shown in Sec.~\ref{kzanderson}.
(b) In Ref.~\cite{Sinha2019}, the driven dynamics in the AA model without the disorder term was studied for changing the quasiperiodic potential. Here we change the disorder strength to cross the AA critical point. Comparing these two cases, we find that although the scaling forms of the KZS are similar, the dimensions of the driving rate are different in two cases. Combining these results, we find that the KZS can apply in the localization transitions for different driving dynamics.

\section{\label{secSum}Summary}
In summary, we have studied the driven dynamics in the localization transitions in $1$D disordered AA model.
By changing the disorder coefficient to cross the critical point, we calculate the dynamics of the localization length $\xi$ and the IPR. For both the critical point of the Anderson model and the AA model, we have verified that the KZS is applicable in characterizing the driven dynamics. Moreover, we have also generalized the KZS to describe the driven dynamics from the excited states. In addition, in the overlap critical region near the AA critical point, we have found that the driven dynamics can be simultaneously described by the KZS with both the critical exponents of the AA model and the critical exponents of the Anderson model. As one possible generalization, one can also investigate the driven dynamics in the many-body localization transition~\cite{Wang2021,Strkalj2021,mblfirst,Xu2019,Mastropietro2015,huseprl,zhangsx2018,Huserev,Altmanrev}.

\section*{Acknowledgments}
B. X. and S. Y. is supported by the National Natural science Foundation of China (Grant No. 12075324), the Science and Technology Projects in Guangzhou (Grant No. 202102020367) and the Fundamental Research Funds for Central Universities, Sun Yat-Sen University (Grant No. 22qntd3005).
L.-J. Zhai is supported by the National Natural science Foundation of China (Grant No. 11704161) and China
Postdoctoral Science Foundation (Grant No. 2021M691535), and Zhongwu Youth Innovation Talent Support Plan of Jiangsu University of technology.

\bibliographystyle{apsrev4-1}

\end{document}